\documentclass[11pt]{article}
\usepackage[affil-it]{authblk}
\usepackage{amssymb} 
\usepackage{amsfonts}
\usepackage{amsmath}
\usepackage{geometry}
\usepackage{graphicx}
\usepackage{lipsum}
\usepackage{hyperref}
\usepackage{color}
\newcommand{\al}{\alpha}
\newcommand{\bt}{\beta}

\newcommand{\ben}{\begin{eqnarray}}
\newcommand{\een}{\end{eqnarray}}
\newcommand{\be}{\begin{equation}}
\newcommand{\ee}{\end{equation}}
\newcommand{\ba}{\begin{eqnarray}}
\newcommand{\ea}{\end{eqnarray}}
\newcommand{\bn}{\begin{equation}\label}

\newcommand{\ga}{\gamma}

\newcommand{\ro}{\rho}

\begin{document}

\title{On sign-changeable interaction in FLRW cosmology}
\author{Fabiola Arevalo}
\affil{N\'ucleo de Matem\'atica, F\'isica y Estad\'istica, Universidad Mayor, Temuco, Chile.\thanks{Electronic address: \texttt{fabiola.arevalo@umayor.cl}}}
\author{Antonella Cid}
\affil{Departamento de F\'isica, Grupo Cosmolog\'ia y Part\'iculas Elementales, Universidad del B\'io-B\'io, Casilla 5-C, Concepci\'on, Chile.\thanks{Electronic address: \texttt{acidm@ubiobio.cl}}}
\author{Luis P. Chimento}
\affil{Departamento de F\'isica, Facultad de Ciencias Exactas y Naturales, Universidad de Buenos Aires e IFIBA, CONICET, Ciudad Universitaria, Pabell\'on I, Buenos Aires 1428, Argentina.\thanks{Electronic address: \texttt{chimento@df.uba.ar}}}
\author{Patricio Mella}
\affil{Instituto de Ciencias F\'isicas y Matem\'aticas, Universidad Austral de Chile, Casilla 567, Valdivia, Chile.\thanks{Electronic address: \texttt{patricio.mella@uach.cl}}}
\date{\today}
\maketitle
  
\begin{abstract}
We investigate an interacting two-fluid model in a spatially flat Friedmann-Lema\^itre-Robertson-Walker (FLRW) Universe, when the energy transfer between these two dark components is produced by a factorisable nonlinear sign-changeable interaction depending linearly on the energy density and quadratically on the deceleration parameter. We solve the source equation and obtain the effective energy densities of the dark sector and their components. We show that the effective equation of state of the dark sector includes some of the several kind of Chaplygin gas equations of state as well as a generalization of the polytropic equation of state. We use bayesian statistics methods to constrain free parameters in the models during its most recent evolution considering supernovae type Ia and measurements of the Hubble expansion rate. The resulting constraints provide new information on sign-changeable interactions, its equivalences and compatibility with previous models and novel late time universe dynamics.
\end{abstract}

\section{Introduction}\label{um}
Modern cosmological observations indicate that the Universe appears to be in a period of accelerated expansion, first noted in Ref.\cite{Riess:1998cb} and confirmed by latest observations \cite{Komatsu:2010fb}. A plethora of theoretical models have been proposed, in which the present acceleration has been modeled by the so-called dark energy component with a characteristic negative pressure that induces late time acceleration in Friedman-Lema\^itre-Robertson-Walker models. 
In addition, it was shown that the cosmological interaction between dark matter and dark energy could address the  late time acceleration of the universe and alleviate the coincidence problem presented in the standard cosmological scenario \cite{coin}. In the literature there are several studies on cosmological interacting models \cite{int,us2,us}, for a general review summarizing these works see \cite{rev}. An interesting aspect of some interacting scenarios consists of alleviating the recently observed tension between high- and low-redshift measurements, namely, the Hubble parameter tension \cite{DiValentino:2017iww}, as well as the $\sigma_8$-tension in the large scale structure formation data \cite{Murgia:2016dug}.

On the other hand, the authors of Ref.\cite{Cai:2009ht} use observational data to analyze a generic type of cosmological interaction {and} find that  {its sign changes} during the evolution of the universe. Later, motivated by Ref.\cite{Cai:2009ht}, the authors of Ref.\cite{Li:2011ga} find a sign change in the cosmological interaction which is described by a running coupling in the cosmic interaction between dark energy and dark matter. Ref.\cite{Sun:2010vz} proposes a model consistent with thermodynamics and observational constraints, where interaction is proportional to the difference between the energy densities of dark components. In the context of this model, there is a natural change in the interaction sign which coincides with the time when dark energy starts to dominate over dark matter during evolution. This change in sign is explored further in \cite{Guo:2017deu} where they analyze a parametrization of the cosmological interaction that changes sign as the scale factor evolves.

In Ref.\cite{Wei:2010cs} a dynamical system analysis was performed for a type of cosmological interaction proportional to the deceleration parameter with the dark energy component modeled by a scalar field, a sign change was naturally induced in the interaction term when the sign of the deceleration parameter changes in the transition from a decelerated universe to an accelerated one, finally the authors find that some scaling attractors could alleviate the cosmological coincidence problem. Other examples of cosmological interaction proportional to the deceleration parameter are found in the literature \cite{5q,Wei:2010fz,Forte:2013fua,Xu:2013iw}.

In the present paper we consider an interacting two-fluid model in which the two dark components  are described as perfect fluids with constant barotropic  indexes and coupled with a nonlinear sign-changeable interaction, depending quadratically on the deceleration parameter. The outline of the paper is as follows. In section \ref{FE}, we propose an interacting two-fluid model and assume a phenomenological cosmological nonlinear sign-changeable interaction between their components. 
In section \ref{qinteraction} we solve exactly the source equation following the procedure outlined in Ref.\cite{Chimento:2009hj} for linear and nonlinear interactions and obtain the total energy densities of the dark sector and their dark components as well as the effective equation of state. In section \ref{data} we obtain the late time behavior of models with cosmological interest 
by using observational data from supernovae type Ia along with measurements of the Hubble expansion rate. Finally, the concluding remarks of this work are presented in section \ref{discussion} for a large set of factorisable nonlinear sign-changeable interactions.

\section{Interacting dark sector with sign-changeable interaction}\label{FE}

Cosmological interaction was mainly introduced to address the late accelerated expansion of the universe, as well as the cosmic coincidence problem  of the standard cosmological scenario ($\Lambda$CDM) \cite{coin}. In this section we investigate interacting scenarios in which the dark components are coupled through sign-changeable interactions $Q$, which are proportional to a quadratic polynomial on the deceleration parameter $q$. This combination includes most of the sign-changeable interactions investigated in the literature  \cite{Wei:2010cs,5q,Wei:2010fz,Forte:2013fua}.

We consider an interacting dark sector for a spatially flat FLRW universe with the line element
\bn{m}
ds^2=dt^2-a^2(t)\left(dX^2+dY^2+dZ^2\right),
\ee
where $t$ is the cosmic time, $a(t)$ is the scale factor, $H=\dot{a}/a$ is the Hubble expansion rate and a dot denotes derivative with respect to the cosmic time. The dark matter and dark energy have energy densities $\rho_m$ and $\rho_x$ respectively, so that
\begin{eqnarray}
\rho&=&\rho_m+\rho_x,\label{rho}\\
\rho'&=&-\gamma_m\rho_m-\gamma_x\rho_x,\label{rhop}
\end{eqnarray}
where $\rho$ is the total energy density and \eqref{rhop} is the conservation equation, the comma indicates derivative with respect to the (time) variable $\eta=\ln{(a^3/a_0^3)}$ and $a_0$ is some value of reference for the scale factor. For the two components we assume equations of state $p_m=(\gamma_m-1)\rho_m$ and $p_x=(\gamma_x-1)\rho_x$, where both barotropic indexes $\gamma_m$ and $\gamma_x$ will be considered constants and satisfying the inequality $\gamma_x<\gamma_m$ throughout this paper. Solving the algebraic system of equations \eqref{rho} and \eqref{rhop} for $\rho_m$ and $\rho_x$ in terms of $\rho$ and $\rho'$, we have 
\bn{rmrx}
\rho_m=-\frac{\gamma_x\rho+\rho'}{\Delta},\quad\quad
\rho_x=\frac{\gamma_m\rho+\rho'}{\Delta},
\ee
where $\Delta=\gamma_m-\gamma_x$ is the determinant of the algebraic system of equations \eqref{rho} and \eqref{rhop}. By using the Friedmann equation we can write the deceleration parameter as a function of the energy density and its $\eta$-derivative,
\bn{q}
q=-\left( 1+\frac{3}{2}\frac{\rho'}{ \rho}\right).
\ee

We begin introducing a factorisable sign-changeable interaction depending quadratically on the deceleration parameter in the form, $Q(\ro,\ro')=\ro[q_1+q_2q+q_3q^2]$, where $q_1$, $q_2$, $q_3$ are constants. Combining \eqref{q} with the proposed interaction $Q$, we find
\bn{gf}
Q(\ro,\ro')=\rho\left(s_1+s_2\frac{\rho'}{\ro}+s_3\left[\frac{\rho'}{\rho}\right]^2\right),
\ee
where the constants $s_i$ are redefinitions of $q_i$. In the first column of table \ref{tableI}, we show five sign-changeable interactions investigated in the literature that result to depend quadratically on the deceleration parameter. At first sight the interactions in the first column of table \ref{tableI} appear to be proportional to the deceleration parameter only, however when one combines \eqref{rmrx} and \eqref{q}, we find that the functions $\ro$, $\rho'$, $\rho_m$, $\rho_x$ and $q$ are related among them. Thus, in the second column of table \ref{tableI} we have written these interactions in terms of $\rho$ and $\rho'$, following the form  given in \eqref{gf}. 
In table \ref{tableI} we show a set of interactions $Q_{1-4}$ investigated in the literature which are included in the general interaction $Q$. In what follows, we will examine the cosmological consequences of the interaction $Q$ in an accelerated scenario for the universe by using the exact dark sector energy density.

\begin{table}[!ht]\centering
\begin{tabular}{c l} \hline
\multicolumn{1}{c}{Interactions} &  \multicolumn{1}{l}{$Q({\rho},{\rho'})$}\\ \hline \hline
$\al_1\, \rho \,q $&$Q_1=-\al_1\,\ro\left( 1+\frac{3}{2}\frac{\rho'}{\ro}\right)$ \\ \hline
$\al_2\, \rho'\,q $&$Q_{2}=-\al_2\,\ro\left(\frac{\rho'}{\ro}+\frac{3}{2}\left[\frac{\rho'}{ \rho}\right]^{2}\right)$  \\ \hline
$\al_3\,\rho_m\,q$ & $Q_{3}=\frac{\al_3}{\Delta}\,\ro \left(\gamma_x+\left(1+\frac{3}{2}\gamma_x\right)\frac{\rho'}{\ro}+\frac{3}{2}\left[\frac{\rho'}{\rho}\right]^{2}\right)$ \\ \hline
$\al_4\,\rho_x\,q$ & $Q_{4}=-\frac{\al_4}{\Delta}\,\ro\left(\ga_m+\left(1+\frac{3}{2}\ga_m\right)\frac{\rho'}{\ro}+\frac{3}{2}\left[\frac{\rho'}{\rho}\right]^{2}\right)$ \\ \hline
$(\al\, \rho+\beta\, \rho')q $ &$Q=-\ro\left(\al+\left(\bt+\frac{3}{2}\al \right)\frac{\rho'}{\ro}+\frac{3}{2}\bt\left[\frac{\rho'}{ \rho}\right]^{2}\right)$\\  \hline
\end{tabular}
\caption{\label{tableI} In the left column, we present five sign-changeable interactions investigated in the literature that involve the deceleration parameter $q$ while in the right, we write these interactions in terms of $\rho$ and $\rho'$ according to \eqref{gf}. Here, $\al_i$, $\al$ and $\bt$ are interaction parameters.}	
\end{table}

From comparing the factorisable sign-changeable interaction \eqref{gf} with the set of interactions $Q_1-Q_4$ and $Q$, we find that the coeficient $s_1, s_2$ and $s_3$ satisfy the following constraint
\be
s_3-\frac{3}{2}s_2=-\frac{9}{4}s_1,
\ee
meaning that interactions in Table 1 include no more than two interaction parameters. 

Cosmological scenarios driven by other interactions depending on $q$, such as the interactions presented in Refs.\cite{Forte:2013fua,Bolotin:2013jpa}, include higher order derivatives of $\rho$ such as $\rho''\rho'/\rho$, which is equivalent to consider an interaction proportional to a quadratic polynomial in $q$ with additional terms proportional to $qq'$ and they lead to a source equation which becomes not integrable by the procedure previously used. 

\section{Solvable sign-changeable  interaction scenario}\label{qinteraction}

We split the conservation equation \eqref{rhop} into two coupled first order differential equations 
\begin{eqnarray}
\rho'_m+\gamma_m\rho_m &=& -Q, \label{rm} \\
\rho'_x+\gamma_x\rho_x &=& Q, \label{rx}
\end{eqnarray}
where we have introduced the phenomenological interaction $Q$ that generates the exchange of energy between dark matter and dark energy components, for $Q>0$ the energy transfer is from dark matter to dark energy and for $Q<0$ we have
an energy transfer from dark energy to dark matter. By differentiating any of equations \eqref{rmrx} and combining with \eqref{rm} or \eqref{rx}, we obtain the second order differential equation that determines the energy density for a given interaction $Q$,  
\begin{equation}
\rho''+(\gamma_m+\gamma_x)\rho'+\gamma_m\gamma_x \rho= \Delta Q. \label{Q}
\end{equation}
This last equation was called ``source equation'' in Ref.\cite{Chimento:2009hj} and it was solved for linear and nonlinear interactions; the former includes a linear combination of the dark matter and dark energy densities, their first derivatives, the total energy density, its first and second derivatives. The latter consists of the above linear combination and additionally significant nonlinear term having the form of a rational function of the dark matter and dark energy densities. By inserting the sign-changeable interaction \eqref{gf} into \eqref{Q}, the source equation becomes a nonlinear differential equation for $\rho$,
\begin{equation}
\rho \rho''+b_1\rho \rho'+b_2\rho^{\prime 2}+b_3 \rho^2=0, \label{ecQI}
\end{equation}
where $b_1=\gamma_m+\gamma_x-s_2\Delta$, $b_2=-s_3\Delta$, $b_3=\gamma_m\gamma_x -s_1\Delta$ are the new parameters of the nonlinear interacting model.

To find the general solution of the nonlinear source equation \eqref{ecQI}, we make the change of variable $X=\rho^{1+b_2}$, valid for $b_2 \neq -1$, in \eqref{ecQI} transforming it in a linear differential equation for $X$. It turns into the equation of a forced dissipative $(b_1 > 0)$ or anti-dissipative $(b_1 < 0)$ linear oscillator
\begin{equation}
X''+b_1 X'+b_3(1+b_2)X=0 ,\label{ecX}
\end{equation}
which has the general solution
\begin{eqnarray}
X=c_1 a^{3\lambda_1}+c_2 a^{3\lambda_2}, \label{rhoX}
\end{eqnarray}
where 
\begin{eqnarray}
\label{l12}
\lambda_{1,2}=\frac{-b_1 \pm \sqrt{b_1^2-4b_3(1+b_2)}}{2},
\end{eqnarray}
are the characteristic roots of the linear differential  equation \eqref{ecX}. {From now on, the $c_i$ with $i=1, 2,..., n$ will denote arbitrary integration constants}. Then, the general solution of the source equation \eqref{ecQI} is obtained from a {\it ``nonlinear superposition"} of the two basis solutions of the second order linear differential equation \eqref{ecX}, so that the energy density $\rho$ has the final form
\begin{eqnarray}
\rho(a)=\left(c_1 a^{3\lambda_1}+c_2 a^{3\lambda_2}\right)^{1/(1+b_2)}.\label{X}
\end{eqnarray}

Besides, by inserting the energy density \eqref{X} into \eqref{rmrx}, we find the dark matter and dark energy densities which are given by
\begin{eqnarray}
\label{r1nl}
\rho_{m}&=&-\left[\lambda_2+\gamma_x(1+b_2)]\rho+c_1(\lambda_1-\lambda_2)\,\frac{a^{3\lambda_1}}{\rho^{b_2}}\right]/[(1+b_2)\Delta],\\
\label{r2nl}
\rho_{x}&=&\left[[\lambda_2+\gamma_m(1+b_2)]\rho+c_1(\lambda_1-\lambda_2)\,\frac{a^{3\lambda_1}}{\rho^{b_2}}\right]/[(1+b_2)\Delta].
\end{eqnarray}
Essentially, after having resolved the source equation for the energy density $\ro$, we also have solved the interacting dark sector model for the individual dark matter and dark energy densities. However, we can also interpret this interacting dark sector model as a unified cosmological model with effective energy density $\ro$, effective equation of state (EoS) $p=-\rho-\rho'$ and effective barotropic index $\ga=-\ro'/\ro$, 
\begin{eqnarray}
\label{e1}
p&=&-\left(1+\frac{\lambda_2}{1+b_2}\right)\rho-\frac{c_1(\lambda_1-\lambda_2)a^{3\lambda_1}}{(1+b_2)\rho^{b_2}},\\
\gamma&=&-\frac{1}{1+b_2}\,\frac{\lambda_1+k\lambda_2a^{3(\lambda_2-\lambda_1)}}{1+ka^{3(\lambda_2-\lambda_1)}},\label{gaf}
\end{eqnarray}
where $k={c_2}/{c_1}$. The equation of state \eqref{e1} includes various of the variable modified Chaplygin gas models investigated in Refs.\cite{Guo:2005qy}-\cite{deb}.

Nonlinear interactions allow the possibility of producing finite-time future singularities \cite{Cataldo:2017nck,Chimento:2015gum,Chimento:2015gga}. In fact, for ${\rm sgn}(c_1)\ne {\rm sgn}(c_2)$ the parentheses in the energy density \eqref{X} vanishes when the scale factor takes the finite value
\bn{as}
a_s=\left(\frac{-c_2}{c_1}\right)^{1/(3\sqrt{b_1^2-4b_3(1+b_2)})}
\ee
it means that both the energy density \eqref{X} and the pressure \eqref{e1}  diverge whenever $b_2<-1$. Hence, in the case that $\ro_s=\ro(a_s)$ diverges at $a_s=a(t_s)$ with $t_s<\infty$, a finite-time future singularity could occur at the cosmic time $t_s$.

By evaluating the energy densities \eqref{r1nl} and \eqref{r2nl} at the present time, such that $a_0=1$, and denoting the density parameters as $\Omega_{m0}=\rho_{m0}/3H_0^2$ and $\Omega_{x0}=\rho_{x0}/3H_0^2$, where $H_0=100h\ [\textrm{km/s/Mpc}]$ is the Hubble parameter, we get 
\begin{eqnarray}
\label{rf}
\rho(a)=\frac{3H_0^2}{(1+k)^{1/(1+b_2)}}\left(a^{3\lambda_1}+k a^{3\lambda_2}\right)^{1/(1+b_2)},
\end{eqnarray}
where
\begin{eqnarray}
\label{k}
k=-\frac{\lambda_1+\gamma_0(1+b_2)}{\lambda_2+\gamma_0(1+b_2)},   
\end{eqnarray}
and $\gamma_0=\gamma_m-\Omega_{x0}\Delta$.

From the energy densities \eqref{rmrx}, we also obtain the cosmic coincidence parameter as:
\begin{eqnarray}
r(a)=\frac{\gamma-\gamma_x}{\gamma_m-\gamma},
\end{eqnarray}
showing that $r$ is positive for $\gamma$ ranging between $\gamma_x$ and $\gamma_m$. 
Evaluating the barotropic index \eqref{gaf} in the limit $a \rightarrow \infty$ we get a finite constant value $r_\infty$, 
\begin{eqnarray}
\label{ri}
r_\infty=-\frac{\lambda_1+\gamma_x(1+b_2)}{\lambda_1+\gamma_m(1+b_2)},
\end{eqnarray}
thus the quadratic sign-changeable interactions generate dark sector models that are possible candidates to alleviate the coincidence problem.
 On the other hand, the deceleration parameter \eqref{q} has a finite constant value $q_\infty$  in the limit $a\rightarrow \infty$ given by, 
\begin{eqnarray}
q_\infty=-1-\frac{3\lambda_1}{2(1+b_2)},
\end{eqnarray}
and the models with parameters satisfying the condition $3\lambda_1/2(1+b_2)>-1$ produce a final accelerated stage.

In the particular case $b_3=0$, we find that $\gamma_m\gamma_x=s_1\Delta$, the characteristic roots are $\lambda_1=0$, $\lambda_2=-b_1$ and the energy density \eqref{X} reads 
\begin{eqnarray}
\rho (a)=\left(c_{1}+c_{2} a^{-3 \gamma_1(1+b_2) }\right)^{1/(1+b_2)}, \label{ChG}
\end{eqnarray}
where $\gamma_1=b_1/(1+b_2)$, while the effective pressure for $b_3=0$,
\begin{eqnarray}
\label{pch}
p=(\gamma_1-1)\rho-\frac{c_1\gamma_1}{\rho^{b_2}},
\end{eqnarray}
turns into the equation of state of a modified Chaplygin gas for positive constants $c_1$ and $c_2$ or a modified anti-Chaplygin gas for ${\rm sgn}(c_1)\ne {\rm sgn}(c_2)$. In the case that $c_1>0$, $c_2>0$ and $b_1>0$, the energy density \eqref{ChG} shows that the universe has a final de Sitter stage with an effective cosmological constant given by the limit $\Lambda_{e}\to c_1^{1/(1+b_2)}$, and the model reduces to the modified Chaplygin gas of Ref.\cite{Benaoum:2002zs}. 

For $b_2=0$, which implies $s_3=0$, we have $\lambda_{1,2}=(-b_1 \pm \sqrt{b_1^2-4b_3})/2$ and then the source equation \eqref{ecQI} turns into a linear second order differential equation and the energy density \eqref{X} becomes a linear superposition of two different powers of the scale factor. Note that, for $s_3\ne 0$, the term proportional to $\rho^{\prime 2}/\rho$ in equation \eqref{gf} is non-vanishing, so it gives rise to the nonlinear superposition effects in the general solution \eqref{X}. 

For $b_1=0$ the constant $s_2=(\ga_m+\ga_x)/\Delta$ and the linear term proportional to $\ro'$ in interaction \eqref{gf} cancels with the dissipative/anti-dissipative term in the source equation \eqref{Q}. In this case, the equation \eqref{ecX} describes a conservative linear oscillator, for $b_3(1+b_2)<0$, with characteristic roots $\lambda_1=-\lambda_2=\sqrt{-b_3(1+b_2)}$ and the energy density is given by \eqref{X}.\\

For some particular values of the constants in \eqref{ecX} we obtain different branches of solutions. For instance, a solution is found for the particular choices $b_1=0$, $b_3=0$, and $b_2=-\ga_p$ the linear source equation \eqref{ecX} reduces to $X''=0$ and its general solution is $X=c_3+c_4\eta$, where $c_3$ and $c_4$ are integration constants. Thus, one finds that the total energy density $\ro=X^{1/(1-\ga_p)}$ has a logarithmic dependence with the scale factor
\bn{poly} 
\ro(a)=\left[c_3+c_4\ln{\left(\frac{a}{a_0}\right)^3}\right]^{1/(1-\ga_p)},
\ee
while the corresponding effective equation of state $p=-\ro-\ro'$ is
\bn{polyp}
p=-\ro+K\ro^{\ga_p},
\ee
where $K=c_4/(\ga_p-1)$. The latter equation can be interpreted as a possible generalization of the polytropic equation of state $p=K\ro^{\ga_p}$, where $K$ is a constant and $\ga_p$ is the polytropic index. 

Another branch of solution is presented when $b_2=-1$ and the source equation \eqref{ecQI} reads
\begin{equation}
\rho \rho''+b_1\rho \rho'-\rho^{\prime 2}+b_3 \rho^2=0.
\label{rhob2}
\end{equation}
By making the change of variable $z=\ro'/\ro$ into equation \eqref{rhob2} we get the linear differential equation $z'+b_1 z+ b_3=0$ whose solution leads to
\begin{equation}
\rho (a)=\rho_0 \ a^{-3\frac{b_3}{b_1}} e^{-\frac{c_5}{b_1}a^{-3b_1}}, \label{26}
\end{equation}
where $\rho_0$ and $c_5$ are integration constants. Asymptotically, when $a\rightarrow\infty$ the total energy density tends to zero {or infinity depending on the sign of the constants $b_1$, $b_3$ and $c_5$.} This scenario may allow a positive or negative barotropic index given by
\bn{b2-1}
\ga=\frac{b_3}{b_1}-c_5\,a^{-3b_1}.
\ee

For $b_1=0$ into equation \eqref{rhob2} we obtain the effective energy density,
\begin{eqnarray}
\label{new1}
\rho(a)=\rho_0 a^{3c_6}e^{-b_3\log(a^3)^2/2},
\end{eqnarray}
{where $\rho_0$ and $c_6$ are integration constants, and in the limit $a\rightarrow\infty$ \eqref{new1} tends to zero for $b_3>0$ or tends to infinity for $b_3<0$. For the energy density \eqref{new1} we have,}
\begin{eqnarray}
\gamma=c_6-b_3\log(a^3) \label{ln2}
\end{eqnarray}
which could lead to a positive or negative barotropic index, regardless of the sign of the constants $b_3$ and $c_6$. In the literature there are other examples of logarithmic equations of state as those seen in equations \eqref{poly} and \eqref{ln2}, e.g.,  \cite{Odintsov:2018obx} and \cite{Yang:2018pej}.

With the above analysis we have an overview of the various effects produced by interaction \eqref{gf} on the dark components, these results were obtained from the complete solution of the nonlinear source equation \eqref{ecQI}. \\

Finally, we comment that the initial interaction proportional to a quadratic polynomial on $q$ can be enlarged by including a term proportional to the first $\eta$-derivative of the deceleration parameter $Q_{q'}\propto\ro q'$. This generates an additional term in the sign-changeable interaction \eqref{gf}, so that it takes the new form $Q_{q'}=Q+s_4\ro''$, with $s_4$ an arbitrary constant. However, it is easy to see that the new term $s_4\ro''$ modifies the coefficient of $\ro''$ in the source equation \eqref{Q}, so that $\ro''\to (1-s_4\Delta)\ro''$, then we solve for $\ro''=[Q\Delta-\ga_m\ga_x\ro-(\ga_m+\ga_x)\ro']/s$, where $s=1-s_4\Delta$ and $Q_{q'}$ becomes 
\bn{qprima}
Q_{q'}=\frac{1}{s}\left\{Q-s_4\,\ro\left[\ga_m\ga_x+(\ga_m+\ga_x)\frac{\rho'}{\ro}\right]\right\}.
\ee
The effect of the new term redefines the coefficients in the sign-changeable interaction $Q$, thus we obtain a new $Q_{q'}$ that has the same form of the sign-changeable interaction \eqref{gf}. A particular case of this interaction was studied in Ref. \cite{Forte:2013fua}.

\section{Observational Analysis}\label{data}
In this section the observational analysis of the sign-changeable interactions $Q_1-Q_4$ and $Q$ in table \ref{tableI} is performed. We consider as free parameters $\alpha_i$, $\alpha$ and $\beta$ describing the cosmological interaction besides $h$, $\Omega_{m0}$ and $\gamma_x$. Also, for the sake of simplicity, we consider $\gamma_m=1$ in all the observational analysis.
We use type Ia supernovae (SNe) data from the Joint Light-curve Analysis (JLA) sample \cite{Betoule} with $N=740$ supernovae over the redshift range $0.01<z<1.2$. We obtain the apparent magnitude of a supernovae in the B-band as,
\begin{eqnarray}
m(z,{\bf p})=\mu(z,{\bf p})+M_B-\alpha_{JLA}\times X_1+\beta_{JLA}\times \mathcal{C},
\end{eqnarray}
where $X_1$ is the time stretching of the light curve, $\mathcal{C}$ is the supernova color at its maximum brightness, $M_B$, $\alpha_{JLA}$, $\beta_{JLA}$ are nuisance parameters (see Ref.\cite{us2} for more details), {\bf p} represents the model's parameters and the distance modulus is given by
\begin{eqnarray}
\mu(z,{\bf p})= 5\log_{10}\left[\frac{\ d_L(z,{\bf p})}{10\rm{pc}}\right],
\end{eqnarray}
for the luminosity distance defined as $d_L(z,{\bf p})=(1+z)\ r(z,{\bf p})$ \cite{Weinberg} and, 
\begin{eqnarray}
r(z,{\bf p})= \int_0^z\frac{c\ dz}{H(z,{\bf p})}.
\end{eqnarray}

We assumed a multivariate Gaussian likelihood in the Monte Carlo analyses for the JLA sample as,
\bn{like}
\mathcal{L}_{JLA}({\bf p})=\exp\left(-\chi^2_{JLA}({\bf p})/2\right),
\ee
where
\begin{eqnarray}
\chi^2_{JLA}({\bf p})=(m_B-m(z,{\bf p}))^TC_{JLA}^{-1}(m_B-m(z,{\bf p})),
\end{eqnarray}
and $C_{JLA}$ is the covariance matrix of the $m_B$ measurements, estimated accounting
for statistical and systematic uncertainties (see Ref.\cite{us2} for more details).

Additionally, we consider $51$ measurements of the Hubble expansion rate (see table 1 of Ref. \cite{Hz}) where the data come from the differential age method for passively evolving galaxies or indirectly through BAO measurements. The $\chi^2$ function for the Hubble expansion rate is defined as,
\begin{eqnarray}
\chi^2_{H}=\sum_i\left(\frac{H_{obs,i}-H_{th}(z_i;{\bf p})}{\sigma_{H}(z_i)}\right)^2,
\end{eqnarray}
where the subscripts $obs$ and $th$ mean observational values and theoretical predictions respectively, $\sigma_H$ denotes the measurement's error and the likelihood is defined as in \eqref{like}.

From here on we use $H(z)=\sqrt{\rho/3}$, where $\rho$ is given by \eqref{rf} rewritten in terms of the cosmological redshift $z$. The constants $\lambda_{1}$, $\lambda_2$, $b_2$ and $k$ in equation \eqref{rf} are rewritten in terms of the model's parameters \textbf{p}=($h$, \{$\alpha_i$, $\alpha$, $\beta$\}, $\Omega_{m0}$, $\gamma_x$). 

In order to find the best fit parameters, we use the Python interface for the \textsc{MultiNest} algorithm \cite{Multinest}, where we assume the following uniform priors: $\alpha_i\in[-1,1]$, $\beta\in[-1,1]$, $\Omega_{m0}\in [0,1]$ and $\gamma_x\in[-1,1]$. In order to break the degeneracy in the $h$ parameter we consider the joint analyses JLA+$H_0$ and JLA+$H_0$+$H(z)$, where $H_0$ is the local measurement for the Hubble parameter \cite{Riess2016}.

The best fit parameters and 1$\sigma$  error are shown in tables \ref{TR1} - \ref{TR4} where we have considered interactions defined in table \ref{tableI}. In order to focus on the interaction parameters we firstly use $\gamma_x=0$ as prior in tables \ref{TR1} and \ref{TR2}, where we also include the fit of the $\Lambda$CDM model. In tables \ref{TR3} and \ref{TR4} we enlarge the parameter space by including $\gamma_x$ as a free parameter. In these tables we include the fit of the $\omega$CDM model as comparison instead of $\Lambda$CDM, given that the former considers the state parameter of the dark energy $\omega$ as a free parameter. In tables \ref{TR1} and \ref{TR3} we show the analysis of the interactions in table \ref{tableI} considering JLA+$H_0$ and in tables \ref{TR2} and \ref{TR4} we expand the data set to include $H(z)$.

\begin{table}[ht!]
\centering
\renewcommand{\arraystretch}{1.3}
\begin{tabular}{c l l c l c} 
\hline
Scenario &\multicolumn{1}{c}{$h$}&\multicolumn{1}{c}{ $\alpha_i$ or $\alpha$} & $\beta$ & \multicolumn{1}{c}{$\Omega_{m0}$} &  $\chi^2_{\textrm{min}}$ \\ \hline  \hline
$\Lambda\rm{CDM}$ &$0.732\pm 0.016$& \multicolumn{1}{c}{-} & -  & $0.298\pm 0.033$ & $682.901$    \\ \hline
$Q_1$ &$0.733\pm 0.016$& $-0.05^{+0.54}_{-0.49}$   & - & $0.286^{+0.13}_{-0.092}$ &  $683.033$   \\ \hline
$Q_{2}$ &$0.733\pm 0.016$& $-0.13^{+0.37}_{-0.78}$ &-  & $0.310^{+0.057}_{-0.050}$ & $682.769$  \\ \hline
$Q_{3}$ &$0.732\pm 0.016$& $0.13^{+0.79}_{-0.37}$  & - & $0.310^{+0.058}_{-0.051}$ & $682.766$ \\ \hline
$Q_{4}$ &$0.732\pm 0.016$& $0.11^{+0.37}_{-0.63}$ & - & $0.308^{+0.083}_{-0.057}$ &  $682.802$ \\ \hline
$Q$ &$0.733\pm 0.016$& $-0.01^{+0.57}_{-0.44}$& $0.05^{+0.49}_{-0.64}$ &$0.293^{+0.13}_{-0.084}$ & $682.993$  \\ \hline
\end{tabular}
\caption{\label{TR1} Best fit parameters and 1$\sigma$ error for the analysis using JLA$+H_0$. We have considered $\gamma_x=0$ and as a comparison we show the fit for $\Lambda$CDM.}
\end{table}

\begin{table}[ht!]
\centering \renewcommand{\arraystretch}{1.3}
\begin{tabular}{c l l c l c} \hline
Scenario &\multicolumn{1}{c}{$h$}& \multicolumn{1}{c}{ $\alpha_i$ or $\alpha$}  & $\beta$ & \multicolumn{1}{c}{$\Omega_{m0}$} &  $\chi^2_{\textrm{min}}$ \\ \hline\hline
$\Lambda\rm{CDM}$ & $0.7106\pm0.0089$ & \multicolumn{1}{c}{-} & -  & $0.256\pm0.013$ & $711.966$ \\ \hline
$Q_{1}$ & $0.707\pm 0.010 $ &  $-0.44^{+0.22}_{-0.36}$ & - & $0.132^{+0.054}_{-0.11} $ & $711.259$ \\ \hline
$Q_{2}$ & $0.7076\pm 0.0098$ & $0.54^{+0.42}_{-0.15}$& -  & $0.218^{+0.023}_{-0.033}$ & $709.588$ \\\hline
$Q_{3}$ &$0.707\pm 0.010$ & $-0.54^{+0.14}_{-0.43}$ & - & $0.219^{+0.022}_{-0.034}$ & $709.508$ \\ \hline
$Q_{4}$ & $0.709^{+0.010}_{-0.011}$ & $-0.01\pm 0.31$ & - & $0.249^{+0.084}_{-0.062}$ & $712.582$ \\ \hline
$Q$ & $0.7063^{+0.0096}_{-0.011}$ & $-0.25\pm 0.34$ & $0.42^{+0.53}_{-0.23}$ & $0.159^{+0.082}_{-0.095}$ & $709.786$ \\ \hline
\end{tabular}
\caption{\label{TR2} Best fit parameters and 1$\sigma$ error for the analysis using JLA$+H_0$+$H(z)$. We have considered  $\gamma_x=0$ and as a comparison we show the fit for $\Lambda$CDM.}
\end{table}

\begin{table}[ht!]
\centering \renewcommand{\arraystretch}{1.3}
\begin{tabular}{c l l c l l c} \hline 
Scenario &\multicolumn{1}{c}{$h$}& \multicolumn{1}{c}{ $\alpha_i$ or $\alpha$}  & $\beta$ & \multicolumn{1}{c}{$\Omega_{m0}$} & \multicolumn{1}{c}{$\gamma_x$} & $\chi^2_{\textrm{min}}$ \\\hline\hline
$\omega\rm{CDM}$  &$0.732\pm 0.016$& \multicolumn{1}{c}{-} & - & $0.220\pm0.110$ & $0.13^{+0.26}_{-0.12}$ & $683.017$ \\\hline
$Q_{1}$ &$0.733\pm 0.016$& $-0.05\pm 0.46$ & - & $0.238^{+0.094}_{-0.21}$ & $0.07^{+0.32}_{-0.13}$& $683.495$ \\\hline
$Q_{2}$ &$0.732\pm 0.016$& $0.01\pm 0.53$ & - & $0.210\pm0.110$ & $0.14^{+0.24}_{-0.12}$ & $683.409$ \\\hline
$Q_{3}$ &$0.732\pm 0.016$& $0.12^{+0.81}_{-0.39}$ & - &$0.24^{+0.11}_{-0.16}$ & $0.11^{+0.30}_{-0.12}$ & $682.885$ \\\hline
$Q_{4}$ &$0.732\pm 0.016$& $0.02^{+0.29}_{-0.57}$ & - & $0.230^{+0.130}_{-0.160}$ & $0.09^{+0.28}_{-0.14}$ & $684.520$\\\hline 
$Q$ &$0.732\pm 0.016$& $-0.03\pm 0.48$ & $0.01^{+0.54}_{-0.63}$ & $0.234^{+0.094}_{-0.21}$ & $0.09^{+0.30}_{-0.12}$ & $683.548$ \\\hline
\end{tabular}
\caption{\label{TR3} Best fit parameters and 1$\sigma$ error for the analysis using JLA$+H_0$. As a comparison we show the fit for $\omega\rm{CDM}$.}
\end{table}

\begin{table}[ht!]
\centering \renewcommand{\arraystretch}{1.3}
\begin{tabular}{c l l c l l c} \hline 
Scenario & \multicolumn{1}{c}{$h$}& \multicolumn{1}{c}{ $\alpha_i$ or $\alpha$}  & $\beta$ & \multicolumn{1}{c}{$\Omega_{m0}$} & \multicolumn{1}{c}{$\gamma_x$}& $\chi^2_{\textrm{min}}$  \\ \hline  \hline
$\omega\rm{CDM}$ & $ 0.7042^{+0.0098}_{-0.011}$ & \multicolumn{1}{c}{-} & - & $0.251\pm 0.015$ & $0.053^{+0.059}_{-0.053}$  & $710.028$ \\ \hline
$Q_{1}$  &$0.7031^{+0.0096}_{-0.011}$ & $-0.49^{+0.19}_{-0.29}$ & - & $0.103^{+0.037}_{-0.094}$ & $0.090^{+0.079}_{-0.065}$ & $708.426$\\\hline
$Q_{2}$ & $0.7044^{+0.0094}_{-0.011} $  & $0.57^{+0.40}_{-0.14}$ & - & $0.207\pm 0.036$ & $0.039^{+0.084}_{-0.064}$ & $708.559$ \\\hline
$Q_{3}$&$0.7042^{+0.0098}_{-0.011}$& $-0.55^{+0.14}_{-0.43}$ & - &$0.214^{+0.025}_{-0.033}$ & $0.035^{+0.080}_{-0.062}$ & $708.578$ \\\hline
$Q_{4}$ &$0.7047^{+0.0099}_{-0.011}$& $-0.307^{+0.088}_{-0.35}$ & - & $0.166^{+0.052}_{-0.099}$ & $0.097^{+0.080}_{-0.054}$ & $709.747$\\\hline 
$Q$ & $0.7030^{+0.0093}_{-0.011}$ & $-0.37^{+0.27}_{-0.39}$ & $0.22^{+0.69}_{-0.34}$ & $0.124^{+0.050}_{-0.11}$ & $0.077^{+0.095}_{-0.065}$ & $708.344$ \\  \hline
\end{tabular}
\caption{\label{TR4} Best fit parameters and 1$\sigma$ error for the analysis using JLA$+H_0$+$H(z)$. As a comparison we show the fit for $\omega\rm{CDM}$.}
\end{table}

We notice in tables \ref{TR1} and \ref{TR3} that the values of the interaction parameters, $\alpha_i$, $\alpha$ and $\beta$, are all consistent with zero (within the 1$\sigma$ region). Nevertheless, only scenarios $Q_4$ and $Q$ are consistent with no interaction in table \ref{TR2} and only scenario $Q$ in table \ref{TR4}. This means that it is necessary to add the $H(z)$ data in order to get a defined sign of the interaction in most of the scenarios. On the other hand, only interaction $Q_1$ and $Q$ maintain the same sign of interaction parameters for all the results in tables \ref{TR1} - \ref{TR4}.

The results in tables \ref{TR3} and \ref{TR4} show a positive $\gamma_x$ in any case, nevertheless this value is consistent with zero in some cases. In comparing tables \ref{TR1} with \ref{TR2} and tables \ref{TR3} with \ref{TR4} for interacting scenarios, we can see that the values of the parameters $h$ and $\Omega_{m0}$ become smaller including the $H(z)$ data. This is natural given that we are using a Gaussian prior to set the value of $h$ in obtaining the results of tables \ref{TR1} and \ref{TR3}, that is, the SNe data do not restrict the $h$ parameter by themselves. In including the $H(z)$ data in tables \ref{TR2} and \ref{TR4} the $h$ posterior changes because this set of data is closely related to the $h$ parameter.

In comparing tables \ref{TR1} with \ref{TR3} and \ref{TR2} with \ref{TR4} we notice that in including the $\gamma_x$ parameter the value of $h$ is approximately the same, nevertheless, the $\Omega_{m0}$ parameter always diminishes. On the other hand, in comparing the full analysis in tables \ref{TR2} and \ref{TR4}, we observe that the sign of the interaction is maintained in each case.
We additionally notice that the fit for interactions $Q_{2}$ and $Q_{3}$ are very similar in tables \ref{TR1} and \ref{TR2}, this is consistent with the fact that both correspond to the same interaction up to a sign in the case $\gamma_x=0$ and $\gamma_m=1$ as we can observe from table \ref{tableI}.

\begin{figure}[ht!]
\centering
\includegraphics[width=0.49\linewidth]{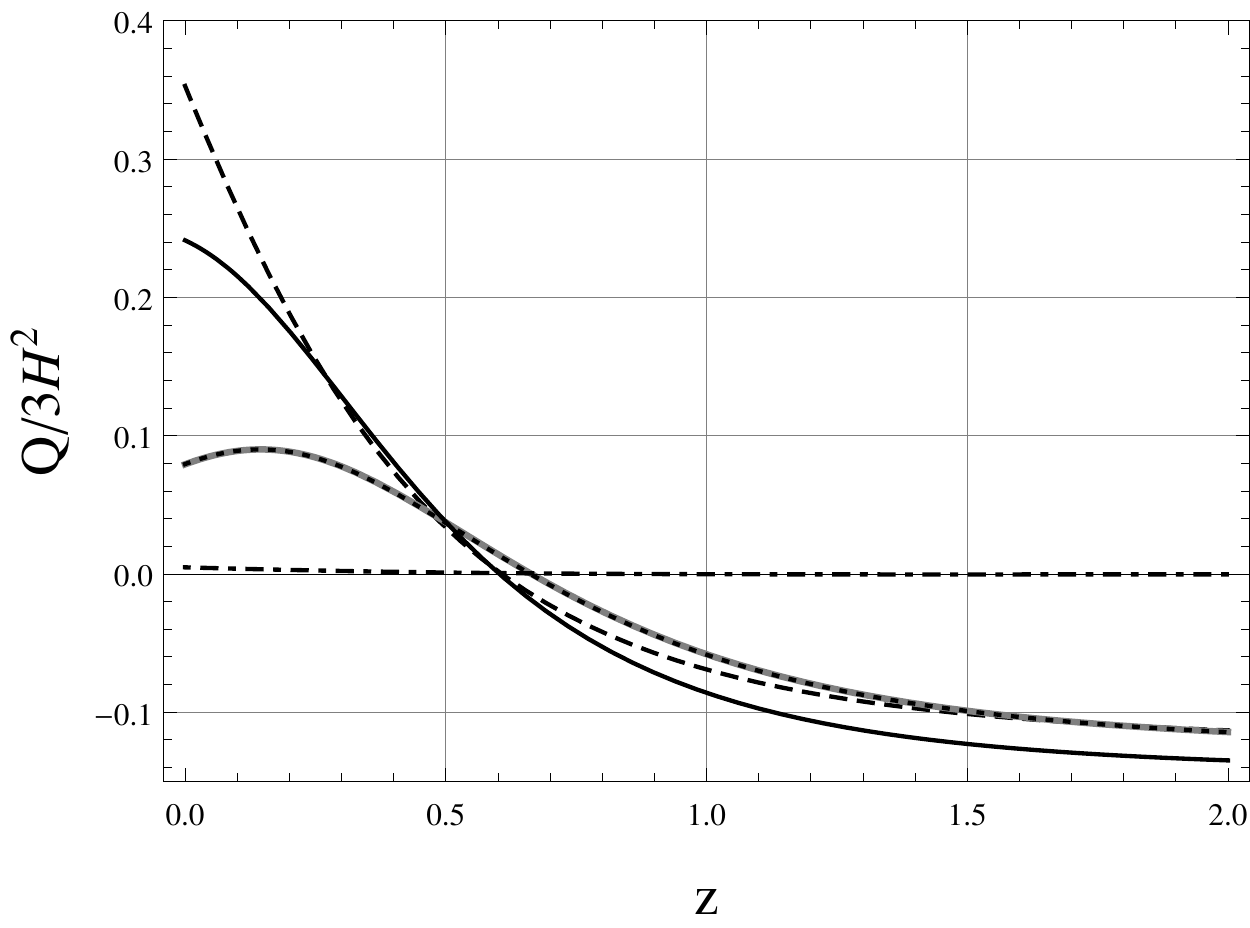}
\includegraphics[width=0.49\linewidth]{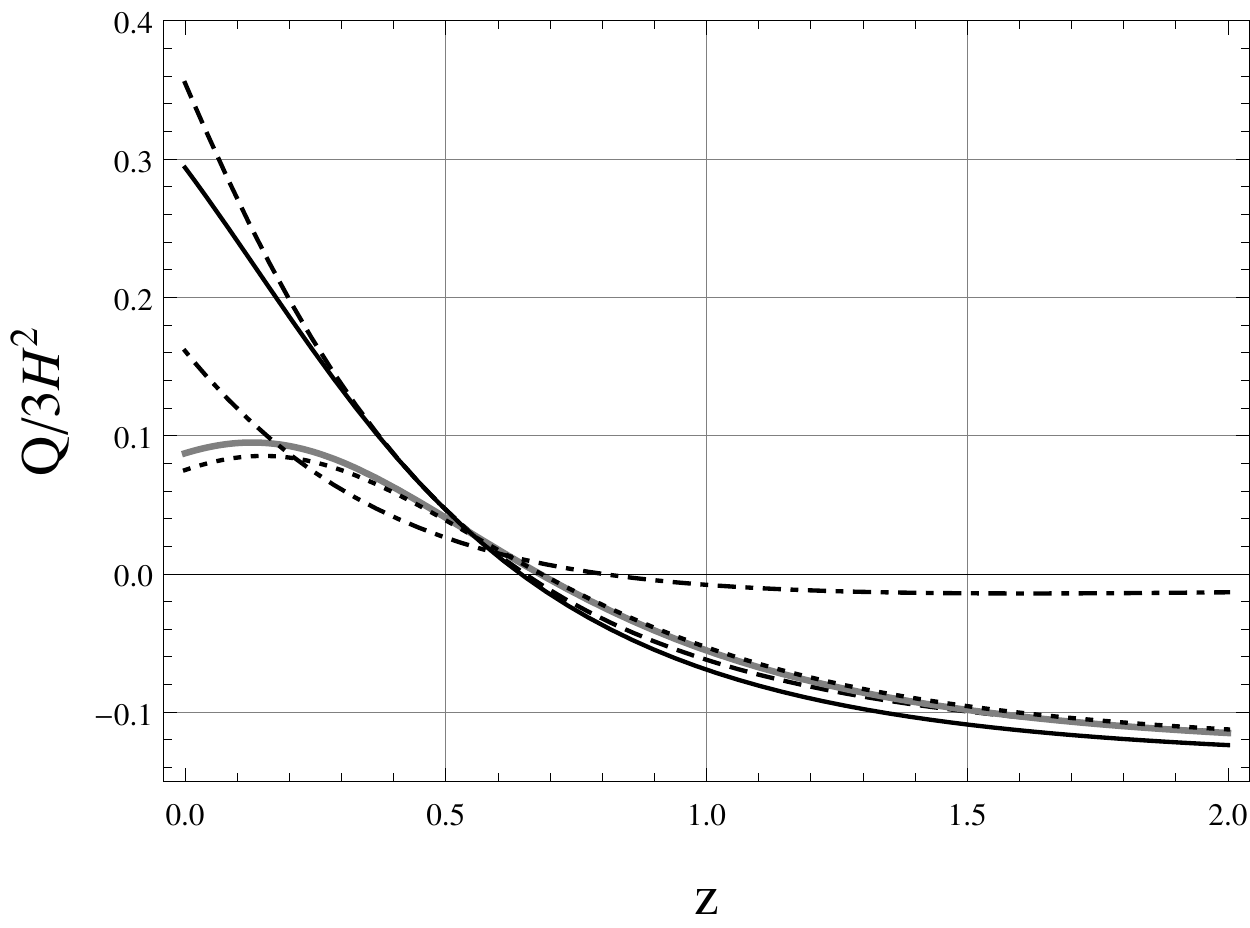}
\caption{\label{fig4}Evolution of a dimensionless interaction term. The left panel corresponds to $\gamma_x=0$ and the right panel to a fitted $\gamma_x$, both figures consider the analysis JLA+$H_0$+H(z). The black lines are for interacting models, dashed, dotted, dot-dashed and solid lines represent interactions $Q_1$ and $Q_3$, $Q_4$ and $Q$, respectively. The gray line is for $Q_2$.}
\end{figure}

\begin{figure}[ht!]
\centering
\includegraphics[width=0.49\linewidth]{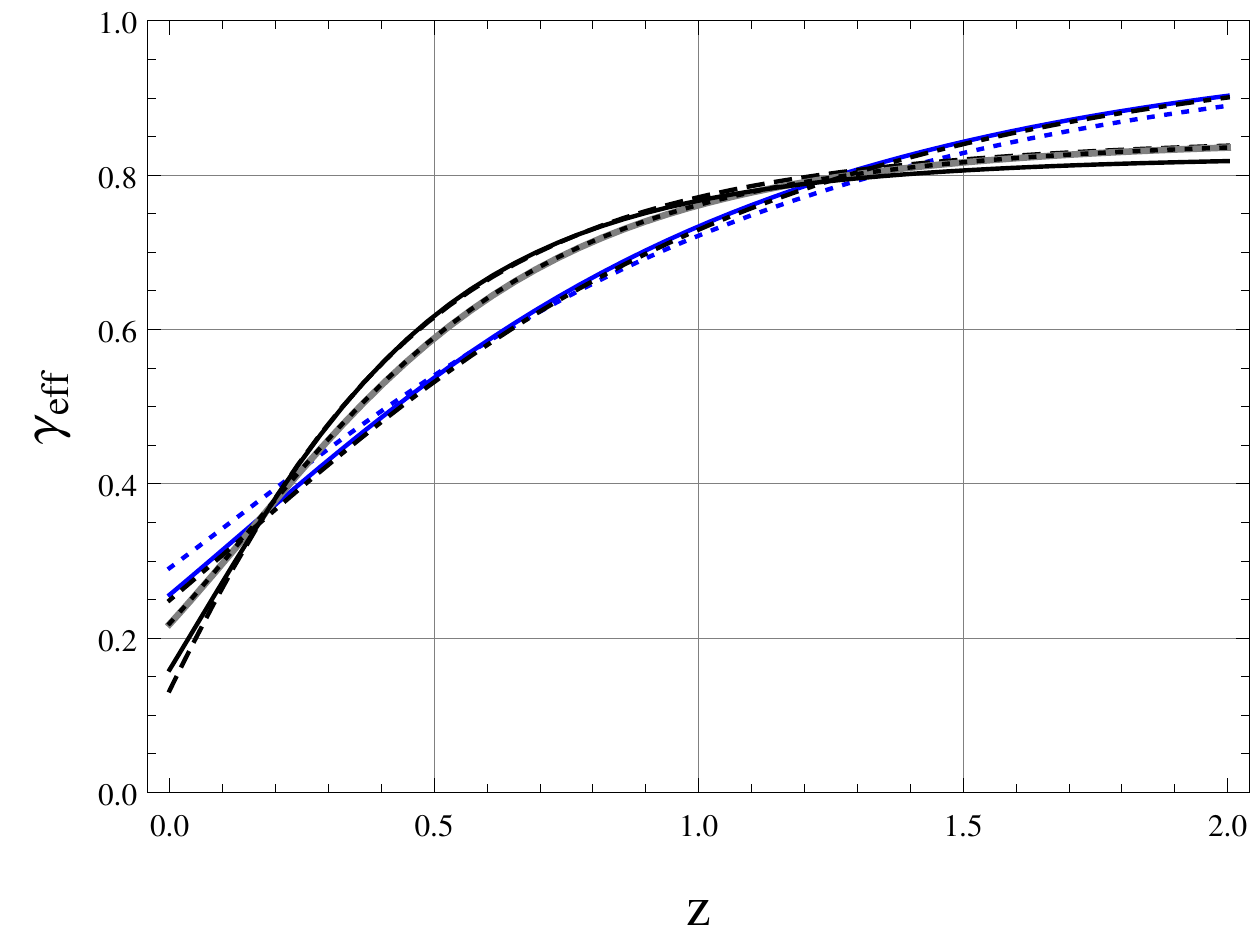}
\includegraphics[width=0.49\linewidth]{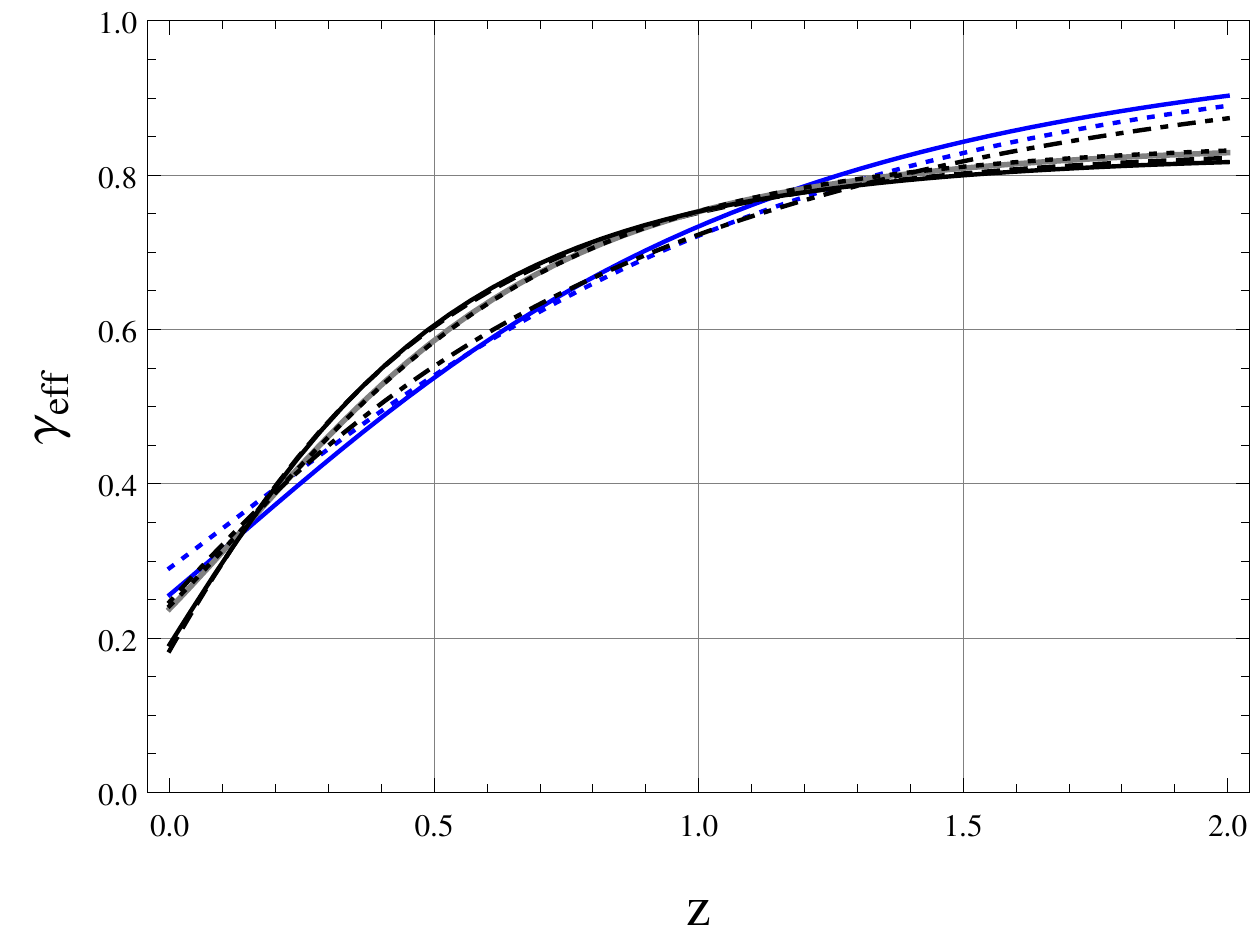}
\caption{\label{fig5}Evolution of the effective barotropic index. The left panel corresponds to $\gamma_x=0$ and the right panel to a fitted $\gamma_x$, both figures consider the analysis JLA+$H_0$+H(z). The blue lines represent $\Lambda$CDM (solid) and $\omega$CDM (dotted). The black lines are for interacting models, dashed, dotted, dot-dashed and solid lines represent interactions $Q_{1}$ and $Q_3$, $Q_4$ and $Q$, respectively. The gray line is for $Q_{2}$.}
\end{figure}

\begin{figure}[ht!]
\centering
\includegraphics[width=0.49\linewidth]{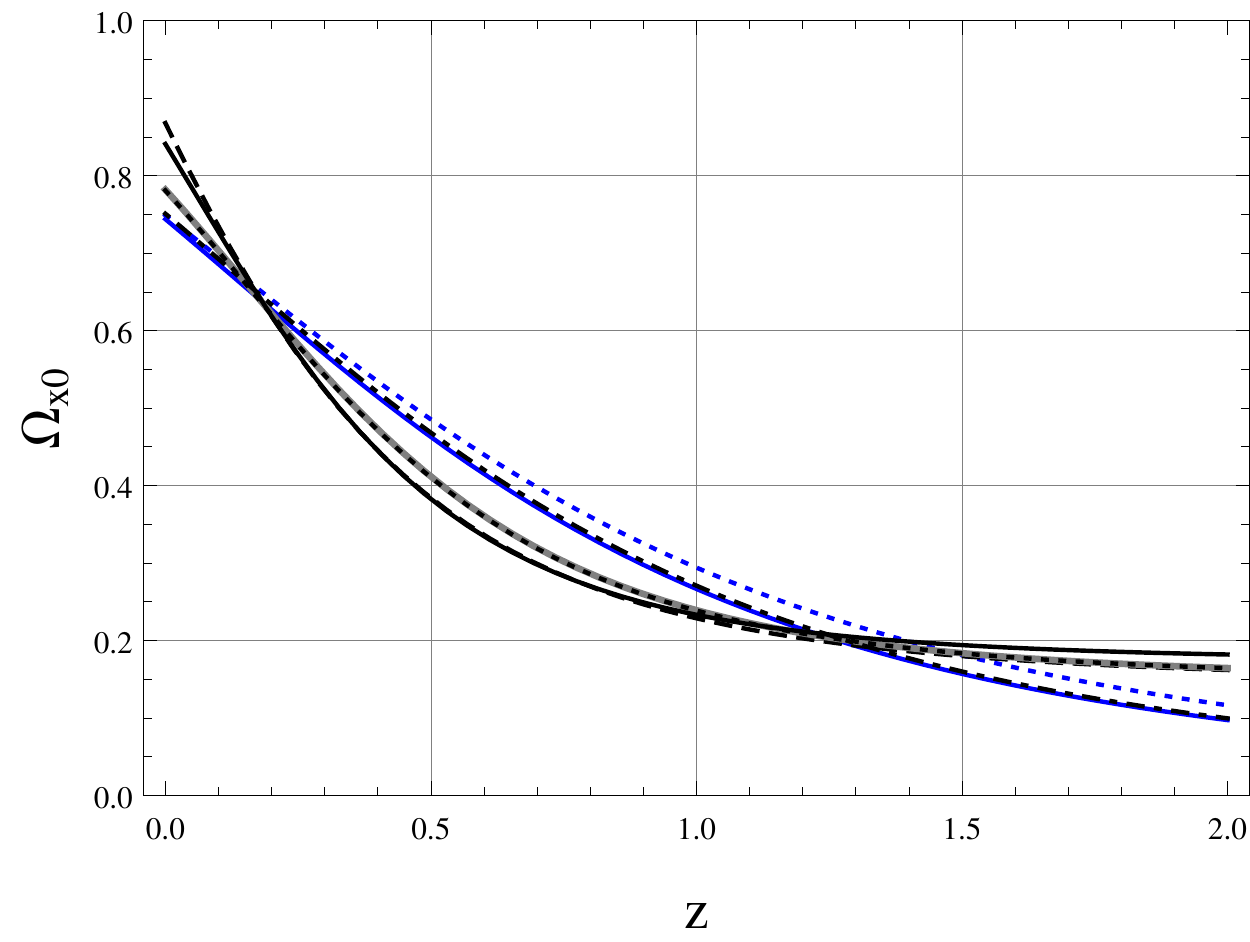}
\includegraphics[width=0.49\linewidth]{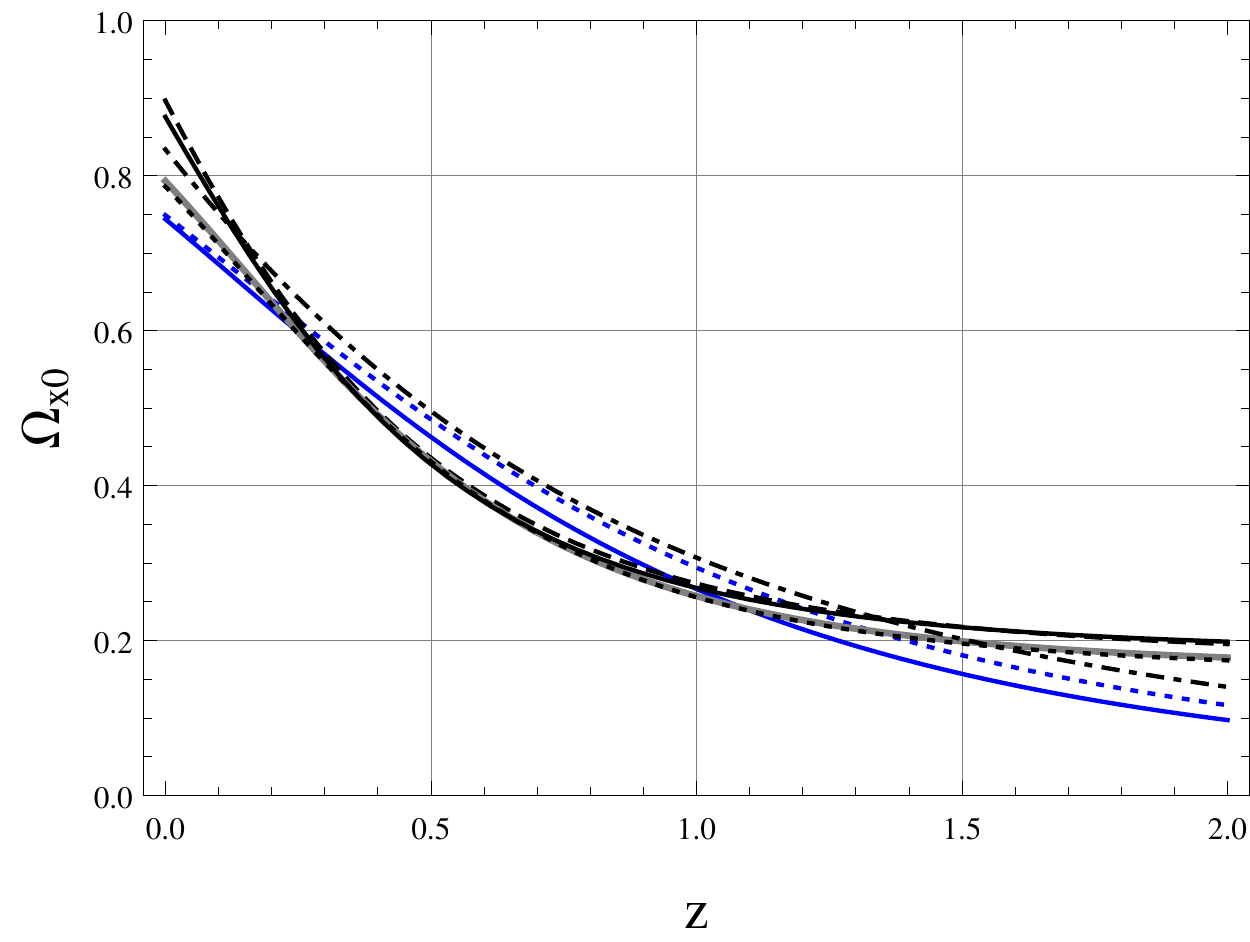}
\caption{\label{fig6}Evolution of the density parameter for dark energy. The left panel corresponds to $\gamma_x=0$ and the right panel to a fitted $\gamma_x$, both figures consider the analysis JLA+$H_0$+H(z). The blue lines represent $\Lambda$CDM (solid) and $\omega$CDM (dotted). The black lines are for interacting models, dashed, dotted, dot-dashed and solid lines represent interactions $Q_{1}$ and $Q_3$, $Q_4$ and $Q$, respectively. The gray line is for $Q_{2}$.}
\end{figure}

\begin{figure}[ht!]
\centering
\includegraphics[width=0.6\linewidth]{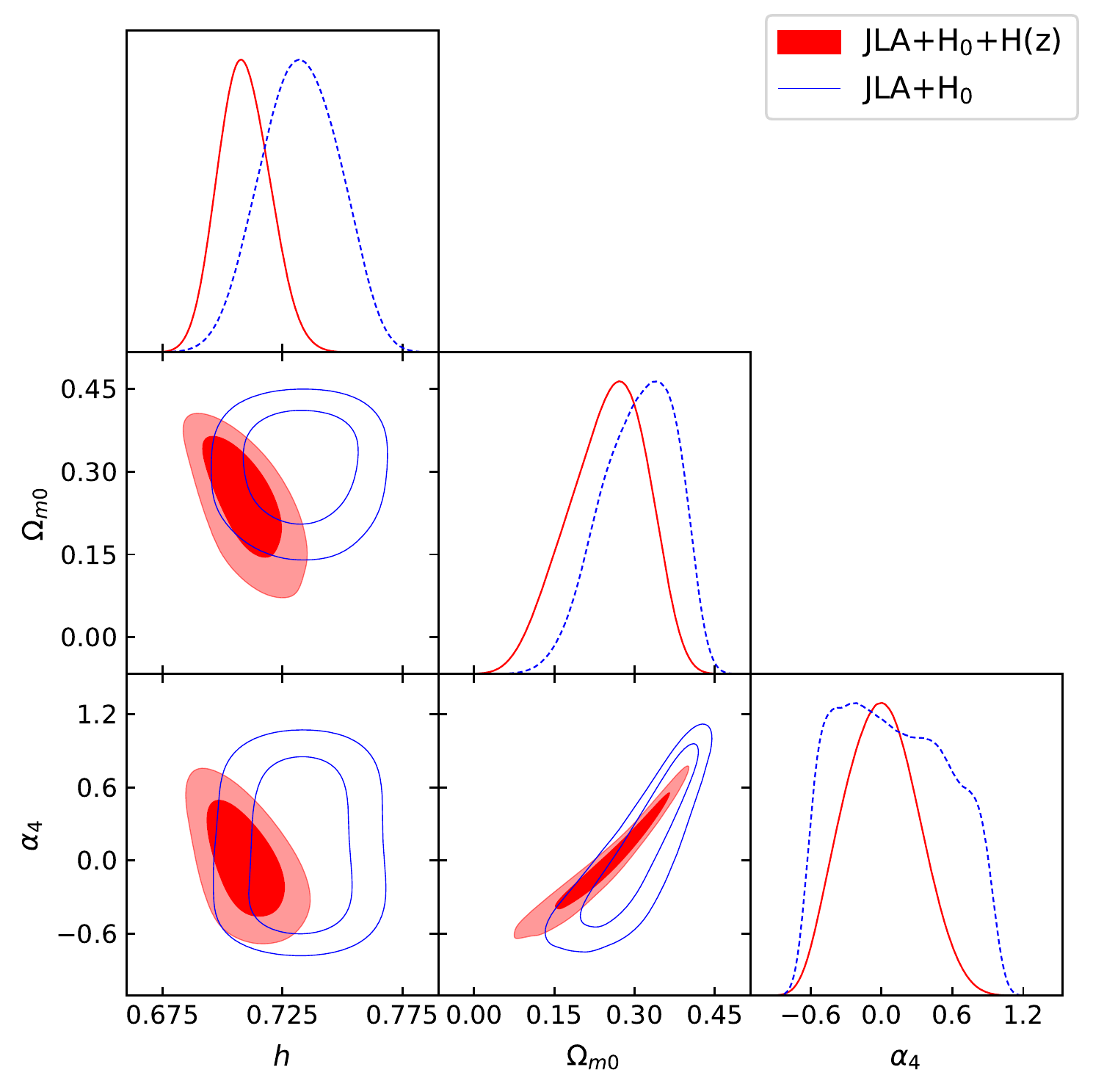}
\caption{\label{fig1}The figure show the contour plot for interaction $Q_4$ with the 1$\sigma$ and 2$\sigma$ regions. We considered the analysis JLA+$H_0$ (blue) and JLA+$H_0$+$H(z)$ (red).}
\end{figure}

\begin{figure}[ht!]
\centering
\includegraphics[width=0.7\linewidth]{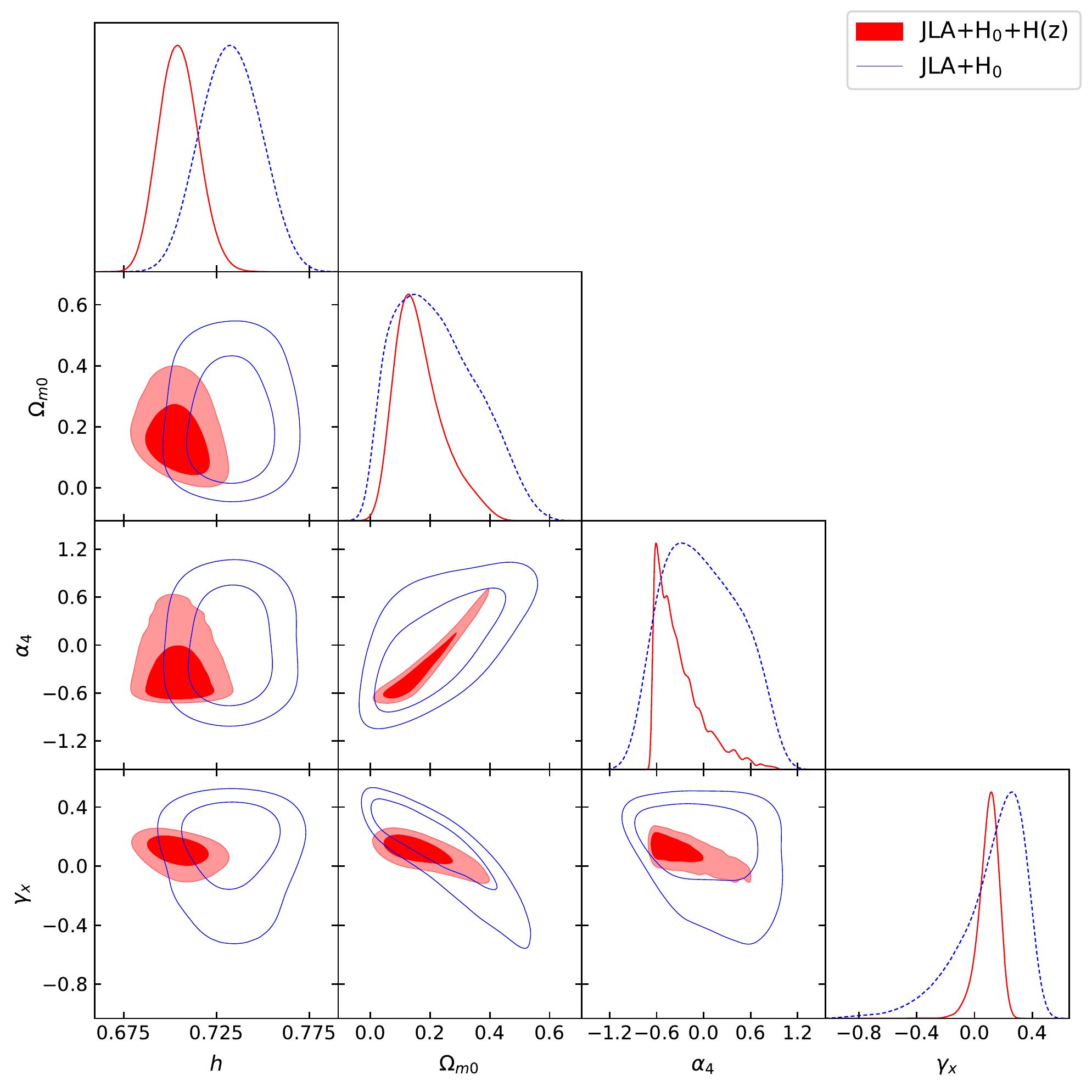} 
\caption{\label{fig2}The figure show the contour plot for interaction $Q_4$ with the 1$\sigma$ and 2$\sigma$ regions. We considered the analysis JLA+$H_0$ (blue) and JLA+$H_0$+$H(z)$ (red).}
\end{figure}

In figure \ref{fig4} we show the evolution of the interaction term and we observe that all the studied interactions, $Q_1-Q_4$ and $Q$, correspond to a positive interacting term today, i.e., we have an energy transfer from dark matter to dark energy today. Besides, the change of sign of the interaction had place in the past (around $z\sim 0.7$), concurrently with the deceleration parameter transition in each case. 

Figure \ref{fig5} shows the effective barotropic index which results to be positive today in each case, in figure \ref{fig6} we show the evolution of the energy density parameter for the dark energy and we notice that the dark energy contribution today is always higher for the studied interacting scenarios, compared to the standard scenario $\Lambda$CDM, this is consistent with an energy transfer from dark matter to dark energy. Furthermore, from figures \ref{fig4}-\ref{fig6} we realize that interaction $Q_4$ (dot-dashed line) corresponds to the sign-changeable scenario closest to the $\Lambda$CDM or the $\omega$CDM evolution.

In figures \ref{fig1} and \ref{fig2} we show as example the contour plot of interaction $Q_4$ in table \ref{tableI}. Figure \ref{fig1} has the $\gamma_x$ parameter set to zero and in figure \ref{fig2} $\gamma_x$ is allowed to vary. We can see that all the parameters are better constrained in the analysis JLA+$H_0$+H(z) compared to the analysis JLA+$H_0$.

{Finally, we notice from the results in tables \ref{TR1} - \ref{TR4} that interacting models $Q_1$-$Q_4$ and $Q$ fulfill the condition $b_2>-1$, hence according to  Eq.\eqref{X}, we do not expect scenarios with a future singularity.}

\pagebreak
\section{Final Discussion}\label{discussion}
We have presented an interacting dark sector scenario in a spatially 
flat FLRW and introduced a factorisable nonlinear sign-changeable interaction between the dark components depending linearly on the energy density and quadratically on the deceleration parameter, $Q =\ro[q_0 +q_1 q +q_2 q^2]$, resulting in cosmological scenarios with a natural sign change related to the accelerated expansion of the universe.

We have studied several linear and nonlinear sign-changeable interactions and written them in terms of $\ro$ and $\ro'$ in table \ref{tableI}, by using the deceleration parameter \ref{gf}. Interactions $Q_1-Q_{4}$ were previously analyzed in a different context in Ref.\cite{Wei:2010cs}, by using dynamical system and considering the cosmological evolution of the dark energy component in the form of quintessence and phantom scalar fields. Instead we have considered a general sign-changeable interaction $Q$ that includes $Q_1-Q_4$ and assumed an interacting dark sector composed of dark matter and dark energy with constant barotropic indexes. We have focused on a larger set of scenarios which can include an inherent change of sign in the interaction term. 

We have shown that different types of sign-changeable interactions are 
linked by recognizing a convenient form of expanding the interaction terms and therefore integration methods described in \cite{Chimento:2009hj} can be applied.
We have solved the corresponding source equation \eqref{Q} when it is sourced by $Q$ depending on the deceleration parameter and obtained the effective energy density \eqref{X}, the dark matter and dark energy densities \eqref{r1nl} and \eqref{r2nl} as well as the effective equation of state \eqref{e1}. In the particular scenario $b_3=0$ in Eq. \ref{ecQI}, we have found that the effective energy density \eqref{ChG} and effective equations of state  \eqref{pch} describe the Chaplygin gas and its generalizations, so we have an unified model for the interacting dark sector.

We have used SNe type Ia from the JLA compilation \cite{Betoule} along with $H(z)$ \cite{Hz} data and the Riess' value for $H_0$ \cite{Riess2016} to constrain the parameters of the five interactions in table \ref{tableI}, using the effective energy density \eqref{X}. The best fit parameters for each model are shown in tables \ref{TR1}-\ref{TR4}, from which we obtain that $b_2>-1$ in any case, then from table \ref{tableI}, we could construct an interacting dark model free of finite-time future singularities, consistent with the cosmological data used.

According to the observational data analysis, summarized in figure 1 for the evolution of the interaction, the energy transfer occurs from dark matter to dark energy today in all scenarios. Also, as expected it is observed a transition in the past where the energy transfer changes direction, this occurs concurrently with the acceleration-deceleration transition. The effective barotropic index results to be of quintessence-type today in any case, see figure \ref{fig5}. Given that the energy transfer is from dark matter to dark energy in all the cases, we see in figure \ref{fig6} that the dark energy density is always larger than the $\Lambda$CDM model today.

In a forthcoming work we will perform a full analysis including baryons and photons in order to include more available data such as baryonic acoustic oscillations and cosmic microwave background measurements.
\\

{\it Acknowledgements.} FA was partially supported from Direcci\'on de Investigaci\'on Universidad de La Frontera, project no. DI17-0075. AC acknowledges the support of Direcci\'on de Investigaci\'on Universidad del B\'io-B\'io through grant no. GI-172309/C. LPC was supported by Universidad de Buenos Aires under Project No. 20020100100147 and CONICET under Project PIP 114-201101-00317.

\end{document}